\documentstyle[12pt]{article}

\skewchar\fivmi='177
\skewchar\sixmi='177
\skewchar\sevmi='177
\skewchar\egtmi='177
\skewchar\ninmi='177
\skewchar\tenmi='177
\skewchar\elvmi='177
\skewchar\twlmi='177
\skewchar\frtnmi='177
\skewchar\svtnmi='177
\skewchar\twtymi='177
\def\@magscale#1{ scaled \magstep #1}
\font\twfvmi  = ammi10   \@magscale5 
\skewchar\twfvmi='177
\skewchar\fivsy='60
\skewchar\sixsy='60
\skewchar\sevsy='60
\skewchar\egtsy='60
\skewchar\ninsy='60
\skewchar\tensy='60
\skewchar\elvsy='60
\skewchar\twlsy='60
\skewchar\frtnsy='60
\skewchar\svtnsy='60
\skewchar\twtysy='60
\font\twfvsy  = amsy10   \@magscale5 
\skewchar\twfvsy='60

\catcode`@=11
\def\un#1{\relax\ifmmode\@@underline#1\else
        $\@@underline{\hbox{#1}}$\relax\fi}
\catcode`@=12

\let\du=\d                      
\let\um=\H                      

\def\a{\alpha}
\def\b{\beta}

\def\d{\delta}
\def\e{\epsilon}

\def\l{\lambda}

\def\s{\sigma}

\def\G{\Gamma}

\font\sc=font005                        
\def\Sc#1{{\hbox{\sc #1}}}      
\font\ooo=circle10                      
\font\ro=manfnt                         
\def\kcl{{\hbox{\ro 6}}}                
\def\kcr{{\hbox{\ro 7}}}                
\def\ktl{{\hbox{\ro \char'134}}}        
\def\ktr{{\hbox{\ro \char'135}}}        
\def\kbl{{\hbox{\ro \char'136}}}        
\def\kbr{{\hbox{\ro \char'137}}}        

\def\ip{{=\!\!\! \mid}}                                    
\def\bo{{\raise.15ex\hbox{\large$\Box$}}}               
\def\pr{\prod}                                          
\def\TH{{\raise.2ex\hbox{$\displaystyle \bigodot$}\mskip-4.7mu \llap H \;}}
\def\face{{\raise.2ex\hbox{$\displaystyle \bigodot$}\mskip-2.2mu \llap {$\ddot
        \smile$}}}                                      

\def\sp#1{{}^{#1}}                              
\def\Tilde#1{{\widetilde{#1}}\hskip 0.03in}                     
\def\Hat#1{\widehat{#1}}                        
\def\Bar#1{\overline{#1}}                       
\def\leftrightarrowfill{$\mathsurround=0pt \mathord\leftarrow \mkern-6mu
        \cleaders\hbox{$\mkern-2mu \mathord- \mkern-2mu$}\hfill
        \mkern-6mu \mathord\rightarrow$}
\def\dvec#1{\vbox{\ialign{##\crcr
        \leftrightarrowfill\crcr\noalign{\kern-1pt\nointerlineskip}
        $\hfil\displaystyle{#1}\hfil$\crcr}}}           
\def\dt#1{{\buildrel {\hbox{\LARGE .}} \over {#1}}}     

\def\frac#1#2{{\textstyle{#1\over\vphantom2\smash{\raise.20ex
        \hbox{$\scriptstyle{#2}$}}}}}                   
\def\ha{\frac12}                                        
\def\sfrac#1#2{{\vphantom1\smash{\lower.5ex\hbox{\small$#1$}}\over
        \vphantom1\smash{\raise.4ex\hbox{\small$#2$}}}} 
\def\bfrac#1#2{{\vphantom1\smash{\lower.5ex\hbox{$#1$}}\over
        \vphantom1\smash{\raise.3ex\hbox{$#2$}}}}       
\def\afrac#1#2{{\vphantom1\smash{\lower.5ex\hbox{$#1$}}\over#2}}    

\newskip\humongous \humongous=0pt plus 1000pt minus 1000pt
\def\caja{\mathsurround=0pt}
\def\eqalign#1{\,\vcenter{\openup2\jot \caja
        \ialign{\strut \hfil$\displaystyle{##}$&$
        \displaystyle{{}##}$\hfil\crcr#1\crcr}}\,}
\newif\ifdtup
\def\panorama{\global\dtuptrue \openup2\jot \caja
        \everycr{\noalign{\ifdtup \global\dtupfalse
        \vskip-\lineskiplimit \vskip\normallineskiplimit
        \else \penalty\interdisplaylinepenalty \fi}}}
\def\li#1{\panorama \tabskip=\humongous                         
        \halign to\displaywidth{\hfil$\displaystyle{##}$
        \tabskip=0pt&$\displaystyle{{}##}$\hfil
        \tabskip=\humongous&\llap{$##$}\tabskip=0pt
        \crcr#1\crcr}}

\def\ref#1{$\sp{#1)}$}

\parskip=\medskipamount                 
\thicklines                         

\def\oldheadpic{                                
        \setlength{\unitlength}{.4mm}
        \thinlines
        \par
        \begin{picture}(349,16)
        \put(325,16){\line(1,0){4}}
        \put(330,16){\line(1,0){4}}
        \put(340,16){\line(1,0){4}}
        \put(335,0){\line(1,0){4}}
        \put(340,0){\line(1,0){4}}
        \put(345,0){\line(1,0){4}}
        \put(329,0){\line(0,1){16}}
        \put(330,0){\line(0,1){16}}
        \put(339,0){\line(0,1){16}}
        \put(340,0){\line(0,1){16}}
        \put(344,0){\line(0,1){16}}
        \put(345,0){\line(0,1){16}}
        \put(329,16){\oval(8,32)[bl]}
        \put(330,16){\oval(8,32)[br]}
        \put(339,0){\oval(8,32)[tl]}
        \put(345,0){\oval(8,32)[tr]}
        \end{picture}
        \par
        \thicklines
        \vskip.2in}
\def\oldtitle#1#2#3#4{\oldheadpic\begin{center}\vglue.5in{\large\bf #1}\\[.6in]
        {#2}\\[.1in] {\it Department of Physics and Astronomy}\\
        {\it University of Maryland, College Park, MD 20742}\\[.6in]
        Physics Publication \#{#3}\\ {#4}\\[1.5in] {\bf Abstract}\\[.1in]
        \end{center} \begin{quotation}}                 
\def\oldTitle#1#2#3#4#5#6#7{\oldheadpic\begin{center} \vglue .4in
        {\large\bf #1}\\[.4in]
        {#2}\\[.1in] {\it Department of Physics and Astronomy}\\
        {\it University of Maryland, College Park, MD 20742}\\[.1in]
        {#3}\\[.1in] {\it {#4}}\\ {\it {#5}}\\[.4in]
        Physics Publication \#{#6}\\ {#7}\\[.5in] {\bf Abstract}\\[.1in]
        \end{center} \begin{quotation}}                 
\def\border{                                            
        \setlength{\unitlength}{1mm}
        \newcount\xco
        \newcount\yco
        \xco=-24
        \yco=12
        \begin{picture}(140,0)
        \put(\xco,\yco){$\ktl$}
        \advance\yco by-1
        {\loop
        \put(\xco,\yco){$\kcl$}
        \advance\yco by-2
        \ifnum\yco>-240
        \repeat
        \put(\xco,\yco){$\kbl$}}
        \xco=158
        \yco=12
        \put(\xco,\yco){$\ktr$}
        \advance\yco by-1
        {\loop
        \put(\xco,\yco){$\kcr$}
        \advance\yco by-2
        \ifnum\yco>-240
        \repeat
        \put(\xco,\yco){$\kbr$}}
        \put(-20,11){\tiny University of Maryland Elementary Particle
Physics University of Maryland Elementary Particle Physics University of
Maryland Elementary Particle Physics}
        \put(-20,-241.5){\tiny University of Maryland Elementary
Particle Physics University of Maryland Elementary Particle Physics
University of Maryland Elementary Particle Physics}
        \end{picture}
        \par\vskip-8mm}
\def\bordero{                                           
        \setlength{\unitlength}{1mm}
        \newcount\xco
        \newcount\yco
        \xco=-24
        \yco=12
        \begin{picture}(140,0)
        \put(\xco,\yco){$\ktl$}
        \advance\yco by-1
        {\loop
        \put(\xco,\yco){$\kcl$}
        \advance\yco by-2
        \ifnum\yco>-240
        \repeat
        \put(\xco,\yco){$\kbl$}}
        \xco=158
        \yco=12
        \put(\xco,\yco){$\ktr$}
        \advance\yco by-1
        {\loop
        \put(\xco,\yco){$\kcr$}
        \advance\yco by-2
        \ifnum\yco>-240
        \repeat
        \put(\xco,\yco){$\kbr$}}
        \put(-20,12){\ooo
bacdefghidfghghdhededbihdgdfdfhhdheidhdhebaaahjhhdahba
hgdedge
   hgfdiehhgdigicba}
        \put(-20,-241.5){\ooo
ababaighefdbfghgeahgdfgafagihdidihiidhiagfedhadbfd
ecdcdfa
   gdcbhaddhbgfchbgfdacfediacbabab}
        \end{picture}
        \par\vskip-8mm}
\def\headpic{                                           
        \indent
        \setlength{\unitlength}{.4mm}
        \thinlines
        \par
        \begin{picture}(29,16)
        \put(165,16){\line(1,0){4}}
        \put(170,16){\line(1,0){4}}
        \put(180,16){\line(1,0){4}}
        \put(175,0){\line(1,0){4}}
        \put(180,0){\line(1,0){4}}
        \put(185,0){\line(1,0){4}}
        \put(169,0){\line(0,1){16}}
        \put(170,0){\line(0,1){16}}
        \put(179,0){\line(0,1){16}}
        \put(180,0){\line(0,1){16}}
        \put(184,0){\line(0,1){16}}
        \put(185,0){\line(0,1){16}}
        \put(169,16){\oval(8,32)[bl]}
        \put(170,16){\oval(8,32)[br]}
        \put(179,0){\oval(8,32)[tl]}
        \put(185,0){\oval(8,32)[tr]}
        \end{picture}
        \par\vskip-6.5mm
        \thicklines}
\def\title#1#2#3#4{\border\headpic {\hbox to\hsize{#4 \hfill UMDEPP #3}}\par
        \begin{center} \vglue .5in {\large\bf #1}\\[.6in]
        {#2}\\[.1in] {\it Department of Physics and Astronomy}\\
        {\it University of Maryland, College Park, MD 20742}\\[1.5in]
        {\bf Abstract}\\[.1in] \end{center} \begin{quotation}}  
\def\Title#1#2#3#4#5#6#7{\border\headpic
        {\hbox to\hsize{#7 \hfill UMDEPP #6}}\par
        \begin{center} \vglue .4in {\large\bf #1}\\[.4in]
        {#2}\\[.1in] {\it Department of Physics and Astronomy}\\
        {\it University of Maryland, College Park, MD 20742}\\[.1in]
        {#3}\\[.1in] {\it {#4}}\\ {\it {#5}}\\[.5in] {\bf Abstract}\\[.1in]
        \end{center} \begin{quotation}}                 
\def\endtitle{\end{quotation}\newpage}                  

\def\sect#1{\bigskip\medskip \goodbreak \noindent{\bf {#1}} \nobreak \medskip}
\def\refs{\sect{References} \footnotesize \frenchspacing \parskip=0pt}
\def\Item{\par\hang\textindent}

\begin{document}

\def\gg{{\hbox{\sc g}}}
\def\nt{$~N=2$~}
\def\gg{{\hbox{\sc g}}}
\def\nt{$~N=2$~}
\def\tr{{\rm tr}}
\def\Tr{{\rm Tr}}
\def\mpl#1#2#3{Mod.~Phys.~Lett.~{\bf A{#1}} (19{#2}) #3}

\def\scst{\scriptstyle}
\def\itrema{$\ddot{\scriptstyle 1}$}
\def\Bo{\bo{\hskip 0.03in}}
\def\lrad#1{ \left( A {\buildrel\leftrightarrow\over D}_{#1} B\right) }

\def\ula{{\underline a}} \def\ulb{{\underline b}} \def\ulc{{\underline c}}
\def\uld{{\underline d}} \def\ule{{\underline e}} \def\ulf{{\underline f}}
\def\ulg{{\underline g}} \def\ulm{{\underline m}}
\def\uln#1{\underline{#1}}
\def\ulp{{\underline p}} \def\ulq{{\underline q}} \def\ulr{{\underline r}}

\def\plpl{{+\!\!\!\!\!{\hskip 0.009in}{\raise -1.0pt\hbox{$_+$}}
{\hskip 0.0008in}}}

\def\mimi{{-\!\!\!\!\!{\hskip 0.009in}{\raise -1.0pt\hbox{$_-$}}
{\hskip 0.0008in}}}

\def\items#1{\\ \item{[#1]}}
\def\ul{\underline}
\def\un{\underline}
\def\-{{\hskip 1.5pt}\hbox{-}}

\def\kd#1#2{\d\du{#1}{#2}}
\def\fracmm#1#2{{{#1}\over{#2}}}
\def\footnotew#1{\footnote{\hsize=6.5in {#1}}}

\def\low#1{{\raise -3pt\hbox{${\hskip 1.0pt}\!_{#1}$}}}

\def\ip{{=\!\!\! \mid}}
\def\unb{{\underline {\bar n}}}
\def\upb{{\underline {\bar p}}}
\def\um{{\underline m}}
\def\up{{\underline p}}
\def\Phib{{\Bar \Phi}}
\def\Phit{{\tilde \Phi}}
\def\Phibt{{\tilde {\Bar \Phi}}}
\def\Db{{\Bar D}_{+}}
\def\gg{{\hbox{\sc g}}}
\def\nt{$~N=2$~}

\border\headpic {\hbox to\hsize{October 1992 \hfill UMDEPP 93--51}}\par
\begin{center}
\vglue .25in

{\large\bf Supersymmetric Soluble Systems Embedded in} \\
{\large\bf Supersymmetric Self--Dual Yang--Mills Theory}$\,$\footnote{This
work is supported in part by NSF grant \# PHY-91-19746.} \\[.1in]

\baselineskip 10pt

\vskip 0.25in

S.~James GATES, Jr.~~and ~Hitoshi NISHINO \\[.2in]
{\it Department of Physics} \\ [.015in]
{\it University of Maryland at College Park}\\ [.015in]
{\it College Park, MD 20742-4111, USA} \\[.1in]
and\\[.1in]
{\it Department of Physics and Astronomy} \\[.015in]
{\it Howard University} \\[.015in]
{\it Washington, D.C. 20059, USA} \\[.18in]

\vskip 1.0in

{\bf Abstract}\\[.1in]
\end{center}

\begin{quotation}

        We perform dimensional reductions of recently constructed self-dual
$~N=2$~ {\it supersymmetric} Yang-Mills theory in $~2+2\-$dimensions into
two-dimensions.  We show that the universal equations obtained in
these dimensional reductions can embed supersymmetric exactly soluble
systems, such as $~N=1$~ and $~N=2$~ supersymmetric Korteweg-de Vries
equations, $~N=1$~ supersymmetric Liouville theory or supersymmetric
Toda theory.
This is the first  supporting evidence for the conjecture that the
$~2+2\-$dimensional self-dual {\it supersymmetric} Yang-Mills theory generates
{\it supersymmetric} soluble systems in lower-dimensions.

\endtitle

\def\doit#1#2{\ifcase#1\or#2\fi}
\def\[{\lfloor{\hskip 0.35pt}\!\!\!\lceil}
\def\]{\rfloor{\hskip 0.35pt}\!\!\!\rceil}
\def\delsl{{{\partial\!\!\! /}}}
\def\caldsl{{\calD\!\!\! /}}
\def\calO{{\cal O}}
\def\asym{({\scriptstyle 1\leftrightarrow \scriptstyle 2})}
\def\Lag{{\cal L}}
\def\du#1#2{_{#1}{}^{#2}}
\def\ud#1#2{^{#1}{}_{#2}}
\def\dud#1#2#3{_{#1}{}^{#2}{}_{#3}}
\def\udu#1#2#3{^{#1}{}_{#2}{}^{#3}}
\def\calD{{\cal D}}
\def\calM{{\cal M}}
\def\tildef{{\tilde f}}
\def\calDsl{{\calD\!\!\!\! /}}

\def\Hat#1{{#1}{\large\raise-0.02pt\hbox{$\!\hskip0.038in\!\!\!\hat{~}$}}}
\def\hati{{\hat{I}}}
\def\dt{$~D=10$~}
\def\alp{\alpha{\hskip 0.007in}'}
\def\oalp#1{\alp^{\hskip 0.007in {#1}}}
\def\naive{{{na${\scriptstyle 1}\!{\dot{}}\!{\dot{}}\,\,$ve}}}
\def\items#1{\vskip 0.05in\Item{[{#1}]}}
\def\item#1{\Item{#1}}

\def\pl#1#2#3{Phys.~Lett.~{\bf {#1}B} (19{#2}) #3}
\def\np#1#2#3{Nucl.~Phys.~{\bf B{#1}} (19{#2}) #3}
\def\prl#1#2#3{Phys.~Rev.~Lett.~{\bf #1} (19{#2}) #3}
\def\pr#1#2#3{Phys.~Rev.~{\bf D{#1}} (19{#2}) #3}
\def\cqg#1#2#3{Class.~and Quant.~Gr.~{\bf {#1}} (19{#2}) #3}
\def\cmp#1#2#3{Comm.~Math.~Phys.~{\bf {#1}} (19{#2}) #3}
\def\jmp#1#2#3{Jour.~Math.~Phys.~{\bf {#1}} (19{#2}) #3}
\def\ap#1#2#3{Ann.~of Phys.~{\bf {#1}} (19{#2}) #3}
\def\prep#1#2#3{Phys.~Rep.~{\bf {#1}C} (19{#2}) #3}
\def\ptp#1#2#3{Prog.~Theor.~Phys.~{\bf {#1}} (19{#2}) #3}
\def\ijmp#1#2#3{Int.~Jour.~Mod.~Phys.~{\bf {#1}} (19{#2}) #3}
\def\nc#1#2#3{Nuovo Cim.~{\bf {#1}} (19{#2}) #3}
\def\ibid#1#2#3{{\it ibid.}~{\bf {#1}} (19{#2}) #3}

\def\szet{{${\scriptstyle \b}$}}
\def\ula{{\un a}}
\def\ulb{{\un b}}
\def\ulc{{\un c}}
\def\uld{{\un d}}
\def\ulA{{\un A}}
\def\ulM{{\underline M}}
\def\cdm{{\Sc D}_{--}}
\def\cdp{{\Sc D}_{++}}
\def\vTheta{\check\Theta}
\def\Pisl{{\Pi\!\!\!\! /}}

\def\fracmm#1#2{{{#1}\over{#2}}}
\def\gg{{\hbox{\sc g}}}
\def\half{{\fracm12}}
\def\ha{\half}

\def\frac#1#2{{\textstyle{#1\over\vphantom2\smash{\raise -.20ex
        \hbox{$\scriptstyle{#2}$}}}}}                   

\def\fracm#1#2{\hbox{\large{${\frac{{#1}}{{#2}}}$}}}

\def\Dot#1{\buildrel{_{_{\hskip 0.01in}\bullet}}\over{#1}}
\def\dt#1{\Dot{#1}}
\def\uln{{\underline n}}
\def\Tilde#1{{\widetilde{#1}}\hskip 0.015in}
\def\Hat#1{\widehat{#1}}

\def\Dot#1{\buildrel{_{_{\hskip 0.01in}\bullet}}\over{#1}}
\def\dt#1{\Dot{#1}}

\oddsidemargin=0.03in
\evensidemargin=0.01in
\hsize=6.5in
\textwidth=6.5in

\noindent 1.{\it ~~Introduction.}~~In recent papers [1-5]
the important connection between the $~N=2$~ superstring and self-dual
Yang-Mills (SDYM) theories has been pointed out, and we have constructed
self-dual {\it supersymmetric} YM (SDSYM) theories in four-dimensional
space-time with the signature $~(+,+,-,-)\,$.\footnotew{We
use the notation $~D=(t,s)$~ for $~D$-dimensional space-time with
$~t$-time and $~s$-space.  When the signature is not important, we also
use the expression $~D=4$, {\it etc}.}~~The importance of
the SDYM theories [6] in such $~D=4$~ space-times is motivated
by the conjecture [7] that {\it all} exactly soluble (bosonic) models in
lower-dimensions can be embedded into the $~D=4$~ SDYM theory.
It is therefore a natural expectation that {\it
supersymmetric} soluble models in lower-dimensions will be generated by
the SDSYM theories in this space-time.

        A typical example of exactly soluble systems is
the Korteweg-de Vries (KdV) equations [8], associated with the affine G-current
(Kac-Moody) algebra, that forms a particular hierarchy structure [8].
The {\it supersymmetrization} of these systems, called supersymmetric KdV
(SKdV) systems, has also been shown to be
possible [9] with $~N=1$~ as well as $~N=2$~ supersymmetries.
The recent works in Ref.~[10] show that obtaining the (bosonic) KdV system
by a dimensional reduction (DR) of the SDYM theory is indeed possible,
by systematically reproducing the hierarchy
structure of the KdV equations, and general embedding algorithms have
also been developed.

        Another typical example of a $~D=2$~ soluble system is
Toda field theory [11], in which the Liouville field equation
is generalized to more general Lie algebras.  It is also possible to
supersymmetrize the Toda field theory based on what is called
contragradient Lie superalgebra with fermionic roots, by
identifying the superspace grading with the Lie superalgebra [12].

In this paper, we take the first step towards the DR of our SDSYM
into $~D=2$, and show how the system
reproduces the sets of equations of these supersymmetric exactly soluble
systems of SKdV and supersymmetric Toda theories.
We give two ways of dimensional reductions of $~D=(2,2)$~ SDSYM to the
$~D=2$.  It is important to note the differences between our two types of
compactifications.  In the {\it first} type of compactification some
components of the original $~D=(2,2)$~ superfield strength are non-zero after
the compactification.  In the {\it second} type {\it all} components of
the superfield strength vanish after the compactification.

\bigskip\bigskip

\noindent2.{\it ~~Dimensional Reduction of the First Type.}~~We
start with the first type of DR of the $~N=2~$ SDSYM [2], and we will see how
the $~D=2,\, N=2$~ SKdV equations are embedded.
The $~D=(2,2),\, N=2$~ SDSYM has the fields $~(A\du\ula I,
\Tilde\l\du{\Dot\a i} I, T^I)$\footnotew{We
adopt the same index notation as in Refs.~[2-5].  For example, for the
$~D=(2,2)$~ vectorial indices we use $~{\scst\ula,~\ulb,~\cdots ~=~
1,~\cdots,~4}$~
and for the chiral (or anti-chiral) spinors we use $~{\scst
\a,~\b,~\cdots~=~1,~2}$~ (or $~{\scst
\Dot\a,~\Dot\b,~\cdots~=~\Dot 1,~\Dot 2})$.} with the field equations
[2-5]

\newpage
$$\li{&F\du{\ula\ulb} I  = \half \e\du{\ula\ulb}{\ulc\uld}F\du{\ulc\uld}
I ~~,
&(2.1) \cr
&i (\s^\ula)\du\a{\Dot\b} \nabla_\ula \Tilde\l\du{\Dot\b i} I = 0 ~~,
&(2.2) \cr
& \Bo T^I - f^{I J K} (\Tilde \l^{i J} \Tilde\l\du i K) = 0~~,
&(2.3) \cr } $$
where $~{\scst i,~j,~\cdots~=~1,~2}$~ are for the ${\bf
2}\-$representation of $~Sp(1)$~ raised and lowered by $~\e^{i j}$~ and
$~\e{\low{i j}}$, while $~{\scst I,~J,~\cdots}$~ are for the adjoint
representation for the Yang-Mills gauge group, which we usually suppress
from now on.  The $~\Tilde\l\du{\Dot\a i } I
$~ is the anti-chiral Majorana-Weyl (MW) spinor gaugino field, and $~T^I~$ is
a real scalar field in the adjoint representation.  The derivative
$~\nabla_\ula~$ is gauge covariant.

        Even without supersymmetry, the possible DR of the system is {\it not}
unique by any means, because there can be many options of appropriate
coordinates, as well as gauge-fixings.  One of the convenient choices [6]
is the following, where the metric tensor is
$$(\eta_{\ula\ulb}) = \pmatrix{0&1&0&0\cr
1&0&0&0\cr 0&0&0&1\cr 0&0&1&0 \cr} ~~.
\eqno(2.4) $$
We call this set of coordinates $~(z,x,y,t)$, and we regard the
coordinates $~(x,t)$ as the final $~D=2$~ dimensions, into which we
perform the DR.  We can now easily see that the self-duality (SD)
condition (2.1) is
$$\li{&F_{x t} = 0~~,
&(2.5) \cr
&F_{y z} = 0~~,
&(2.6) \cr
& F_{z x} = F_{t y} ~~,
&(2.7) \cr } $$
when $~\e^{z x y t} = +1$.
Eq.~(2.5) implies that the YM gauge fields $~A_x$~ and $A_t$~
in $~D=2$~ should be {\it pure gauge}, namely they can be completely
gauged away!  Therefore we can impose the conditions
$$ A_x = A_t= 0 ~~.
\eqno(2.8) $$
If we further require the independence of all the quantities on the $~y$~
and $~z\-$coordinates [13], eq.~(2.6) and (2.7) respectively imply that
$$\li{& \[ P, B \] = 0 ~~,
&(2.9)  \cr
&\Dot P + B' = 0 ~~,
&(2.10)  \cr} $$
where $~ P\equiv A_y,~ B\equiv A_z$, and
their {\it prime} and {\it dot} denote respectively the derivatives
$~\partial_x\equiv \partial/\partial x$~ and $~\partial_t\equiv
\partial/\partial t$.

We next show the significance of the gaugino field equation (2.2).
A convenient representation for the
$~\s\-$matrices is the frame where the $~\G_5$~
is diagonalized [2].  In accord also with our choice of coordinates, we
find the corresponding explicit $~\s\-$matrices
$$\li{&\s^x = \fracm 1{\sqrt2} (\s^1 + \s^4 )
= \pmatrix{+i & -i \cr -i & +i \cr} ~~, ~~~~
\s^z = \fracm 1{\sqrt2} (\s^1 - \s^4 )
= \pmatrix{-i & -i \cr -i & -i \cr} ~~, \cr
&\s^y = \fracm 1{\sqrt2} (\s^2 + \s^3 )
= \pmatrix{+1 & -1 \cr +1 & -1 \cr} ~~, ~~~~
\s^t = \fracm 1{\sqrt2} (\s^2 - \s^3 )
= \pmatrix{-1 & -1 \cr +1 & +1 \cr} ~~,
&(2.11) \cr
&\Tilde \s^x = \fracm 1{\sqrt2} (\Tilde \s^1 + \Tilde \s^4 )
= \pmatrix{+i & +i \cr +i & +i \cr} ~~, ~~~~
\Tilde \s^z = \fracm 1{\sqrt2} (\Tilde \s^1 - \Tilde \s^4 )
= \pmatrix{-i & +i \cr +i & -i \cr} ~~, \cr
&\Tilde \s^y = \fracm 1{\sqrt2} (\Tilde \s^2 + \Tilde \s^3 )
= \pmatrix{-1 & +1 \cr -1 & +1 \cr} ~~, ~~~~
\Tilde \s^t = \fracm 1{\sqrt2} (\Tilde \s^2 - \Tilde \s^3 )
= \pmatrix{+1 & +1 \cr -1 & -1 \cr} ~~,
&(2.12) \cr} $$
where $~\s^1,~\cdots,~\s^4$~ and $~\Tilde \s^1,~\cdots,~\Tilde\s^4$~ are
the $~\s\-$matrices in the metric $~{\rm diag}.\,(+,+,-,-)$~ in the
representation diagonalizing $~\G_5$~ [2,5].  Since the
$~\Tilde\l_{\Dot\a\,i}$~
is a MW spinor, its second component is to be the complex conjugate of
the first one [2].  Therefore, under the identification
$$~(\Tilde\l_{\Dot\a i}) = \fracm 1{\sqrt 2} \pmatrix {\psi_i -i
\chi\low i \cr
\psi_i +i\chi\low i \cr} ~~,
\eqno(2.13) $$
the gaugino field equation (2.2) yields the two equations
$$\li{&\Dot \psi{}_i  = \chi_i{\,}' ~~ ,
& (2.14)  \cr
& \[ P, \chi_i \, \] + \[ B, \psi_i \, \] = 0 ~~.
& (2.15) \cr } $$

        Finally performing the DR for (2.3), we get
$$\[ B , {T^{\,}}' \, \] + \[P ,\Dot T \, \] + \[ \psi^i ,\chi\low i\,
\] = 0 ~~.
\eqno(2.16) $$
Interestingly, all the linear terms vanish, leaving only commutator
terms.  In summary, the full set of field equations is (2.9), (2.10), (2.14) -
(2.16).

        In the usual simple DR [13], where any dependence on the extra
coordinates is suppressed, the original maximal supersymmetry is to be
preserved.  In our system, the starting \hbox{$~N=2$~} supersymmetry is
expected to yield the $~N=(8,8)$~ supersymmetry in $~D=2$.
We see that this is indeed
the case, by inspecting the DR of the supertranslation rules.  Our
$~D=(2,2),\,N=2$~ supertranslation rules are\footnotew{Our
normalization for the gaugino is different from Ref.~[2] to simplify the DR.}
$$\eqalign{& \d A_\ula{} = -i\big(\e^i\s_\ula \Tilde \l_i \big) ~~,
{}~~~~ \d T = (\Tilde\e^{\,i} \Tilde\l_i) ~~, \cr
&\d \Tilde \l_i =  \fracm 1 8  \left(\Tilde\s^\ula\s^\ulb -
\Tilde\s^\ulb \s^\ula \right) \Tilde\e_i F_{\ula\ulb} - i \Tilde \s^\ula
\e_i \nabla_\ula T ~~. \cr
\cr }
\eqno(2.17) $$
Using our $~\s\-$matrices (2.11) and (2.12) in (2.17), we easily
get $~\d P = \d A_y~$ {\it etc.}~as well as $~\d\psi$~ and $~\d\chi$:
$$\eqalign{&\d P = - \sqrt2 (\b^i\psi_i ) ~~,
{}~~~~ \d B = \sqrt2 (\b^i\chi_i) ~~,  \cr
& \d\psi^i = - \Tilde \b^i P' - \Tilde\a^i \Dot P + {\sqrt2} \b^i T' ~~, \cr
& \d\chi^i = \Tilde\a^i \Dot B + \Tilde\b^i B' +{\sqrt2} \b^i \Dot T ~~, \cr
&\d T = - (\Tilde\a^i \chi_i) - (\Tilde\b^i\psi_i) ~~, \cr }
\eqno(2.18) $$
where $~\a$'s and $~\b$'s are defined by
$$\eqalign{&\a_i \equiv \fracm1{\sqrt2} (\e\ud 1 i + \e\ud 2i  )~~, ~~~~
\b_i \equiv -\fracm i {\sqrt2} (\e\ud 1i - \e\ud 2i ) ~~, \cr
&\Tilde\a_i \equiv \fracm1{\sqrt2} (\Tilde\e\ud {\Dot 1}i + \Tilde\e\ud
{\Dot 2}i) ~~,
{}~~~~\Tilde\b_i = -\fracm i{\sqrt2} (\Tilde\e\ud {\Dot 1}i - \Tilde\e\ud
{\Dot 2}i) ~~, \cr }
\eqno(2.19) $$
for the supertranslation parameters $~\e^\a{}_i$~ and $~\Tilde \e^{\Dot
\a}{}_i$.  The $~\a^i\-$parameter is to be put to zero, and it disappears
in the rule (2.18), in accord with our gauge condition (2.8).   This is
because the $~D=4$~ supertranslation gives $~\a^i\-$parameter in $~\d A_x$~ and
$~\d A_t$.  It is not difficult to confirm the consistency of the
above-obtained field equations (2.9) - (2.11), (2.14) - (2.16) under
supersymmetry (2.18).

        We now come to the embedding of $~N=2$~ SKdV equations in the above
DR, as an example of soluble system.  For this purpose, we
start with an {\it Abelian} SDSYM, which is the simplest known embedding for
the $~N=2$~ SKdV equations, as described in Ref.~[14].  Accordingly
all the commutators disappear, simplifying the computations.
Eventually we have found the following identifications are consistent
for the lowest flow of $~N=2$~ SKdV:
$$\eqalign{& P \equiv u ~~, \cr
& B\equiv u '' + 3u^2 + 3 \xi_i\xi_i\, ' +(a+1) w'^{\,2} + (a-2) w w'' - 3a
u w^2 - 3a \e{\low{i j}} w\xi_i \xi_j ~~, \cr
&\psi_i \equiv \xi_i\, ' ~~, \cr
&\chi_i \equiv - \xi_i''' + 3(u\xi_i)' + 3a (w^2\xi_i)' + \e{\low{i j}}
\left[ (a+2) (w\xi_j\, ')' + (a-1) (w'\xi_j)' \right\}  ~~. \cr }
\eqno(2.20) $$
At this stage, the original $~Sp(1)$~ symmetry has been lost, and all the
repeated indices $~{\scst i,~j,~\cdots~=~ 1,~2}$~ are now contracted by
$~\d_{i j}$~ instead of $~\e{\low{i j}}$.

We can easily confirm that eq.~(2.10) yields
$$\eqalign{\Dot u = \, & - u''' + 6u u' - 3 \xi_i\xi_i \, ''
- 3a w' w'' - (a-2) w w''' \cr
& + 3a (u w^2)' + 3a\e{\low{i j}} (w\xi_i\xi_j)' ~~, \cr }
\eqno(2.21) $$
while (2.14) gives
$$\eqalign{\Dot\xi_i{}' = \, & \bigg[- \xi_i\, ''' + 3 (u\xi_i) '
+ 3a (w^2 \xi_i)'  \cr
& + \e{\low{i j}}\left\{ (a+2) (w\xi_j\, ')' + (a-1) (w'\xi_j)' \right\}
\bigg]'  ~~,
\cr}
\eqno(2.22) $$
which has overall extra derivatives of $~\partial_x$, compared with the
standard $~N=2$~ SKdV equation [9].  These extra derivatives can be
avoided by considering the usual boundary conditions
$$ \lim_{|x| \rightarrow \infty} u(x,t) = 0~~, ~~~~
\lim_{|x| \rightarrow \infty} \xi_i(x,t) = 0~~, ~~~~
\lim_{|x| \rightarrow \infty} w(x,t) = 0~~,
\eqno(2.23) $$
at any arbitrary time $~t$.  The general solution to (2.22) has the form
$$\eqalign{\Dot\xi_i = \, & - \xi_i\, ''' + 3 (u\xi_i) ' + 3a (w^2 \xi_i)'
\cr
& + \e{\low{i j}} \Big[ (a+2) (w\xi_j\, ')' + (a-1) (w'\xi_j)' \Big]
+X_i(t) ~~, \cr }
\eqno(2.24) $$
where $~X_i(t)$~ is an arbitrary function {\it only} of $t$.  However, such
a function is excluded by the boundary condition (2.23).
Thus (2.22) {\it uniquely} yields the fermionic equation in the $~N=2$~
SKdV system [9].

The standard $~w\-$equation in the $~N=2$~ SKdV is derived by taking the
supertranslations of the above field equations.  The supertranslation
rules are obtained from (2.18) with
$$\Tilde\a_i \equiv 0 ~~, ~~~~
\b_i \equiv \fracm 1{\sqrt2} \e\low i ~~, ~~~~
\Tilde\b_i \equiv \e\low i ~~,
\eqno(2.25) $$
as
$$\eqalign{&\d w = - \e\low i \xi_i\, ' ~~, ~~~~ \d u = - \e{\low{i j}}
\e\low i \xi_j\, ' ~~, \cr
&\d\xi_i =  \e{\low{i j}} \e\low j u + \e\low i w' ~~, \cr }
\eqno(2.26) $$
in agreement with the usual rule [9].
Originally we had obtained $~(\d\xi_i)' = (\e{\low{i j}} \e{\low j}u
+ \e_j w')'$~
and similarly for $~\d w'$~ with extra overall $~x\-$derivatives.
However, these derivatives are integrated with {\it no} room
for arbitrary $~x\-$independent functions, again
due to the boundary condition (2.23).
These supertranslations are valid {\it on-shell}, namely we have used the
field equations (2.21) and (2.24).  The
$~w\-$equation obtained by applying (2.26) to (2.21) and (2.24) has
overall $~x\-$derivative, which can be excluded
again by our boundary condition (2.23).  Eventually it coincides with the
standard one [9]:
$$\Dot w = - w''' + 6a w^2 w' + (a+2) (u w)' + \half (a-1) \e\low{i j}(\xi_i
\xi_j)' ~~.
\eqno(2.27) $$

        Note that our method here is easily applicable to higher flows in
the $~N=2$~ SKdV, because the patterns of (2.10) and (2.14)
are common to higher flows [9].

        We mention that since the $~N=2$~ SKdV system has the $~N=1$~ SKdV
equations as its sub-case, the latter can be also embedded in the above
system.  This is seen by putting $~w(x,t) = \xi_2(x,t)=0$~ and $~\xi_1
(x,t) \equiv\xi(x,t)$.

\bigskip\bigskip

\noindent 3.{\it~~Dimensional Reduction of the Second Type.}~~We next give the
DR of the second type, which has more manifest supersymmetry,
embedding the $~N=2$~ SKdV
equations [9] and $~N=1$~ supersymmetric Toda theories [12].

        In the starting $~D=(2,2)$~ SDSYM theory, the superfield $~F_{A B}$~
maintains only half of its original components, due to the SD condition [2].
For our purpose of obtaining the supersymmetric Toda theories, we can
truncate even
more components upon the DR.  Actually this is possible consistently, if we
truncate {\it all} of its components in $~D=(2,2)$.
To see this is consistent with our starting $~N=2$~ SDSYM theory,
we review its $~D=4$~ constraints [2-5], which are
$$\eqalign{& F_{\a i\, \ulb}
= - i (\s_\ulb)_{\a\Dot\b} \Tilde\l\ud{\Dot\b} i ~~, \cr
& F_{\Dot \a i\, \ulb} = 0 ~~, ~~~~ F_{\a i \,\Dot\b j} = 0 ~~, ~~~~
F_{\Dot\a i\, \Dot\b j} = 0 ~~, \cr
& F_{\a i\, \b j} = 2 C_{\a\b} \e{\low{i j}} T ~~, \cr
& \Tilde\nabla_{\Dot\a i} T = - \Tilde\l_{\Dot\a i} ~~, ~~~~
\nabla_{\a i} T = 0 ~~, \cr
&\nabla_{\a i}\Tilde\l_{\Dot\b j} = - i \e{\low{i j}} (\s^\ulc)_{\a\Dot\b}
\nabla_\ulc T ~~, \cr
&\Tilde\nabla_{\Dot\a i}\Tilde\l_{\Dot\b j} = \fracm 14\e{\low{i j}}
(\s^{\ulc\uld})_{\Dot\a\Dot\b} F_{\ulc\uld} ~~. \cr }
\eqno(3.1) $$
For the present DR, we impose the condition\footnotew{By this
choice, we are off the original Wess-Zumino gauge, but this
simplifies the system compared with the first type DR.}
$$\Tilde\l_{\Dot\a i} = 0~~,~~~~ T = 0~~,
\eqno(3.2) $$
Accordingly, the equations
$~F_{x y} =F_{x z} =F_{y z} = F_{y t} = F_{z t} = 0$~ are satisfied, and we are
left with the {\it purely} $~D=2$~ superfields $~F_{x t} = 0,~F_{{\ul\a} x} =
0,~ F_{{\ul\a} t} =0$.
This amounts to the requirement of the $~D=2$~ superspace
constraints
$$F_{A B} = 0~~,
\eqno(3.3) $$
with the superspace indices $~{\scst A,~B,~\cdots}$,
satisfying the Bianchi identities in a trivial way:
$$\nabla_{\[ A} F_{B C)} - T\du{\[ A B|} D F_{D|C)} \equiv 0 ~~.
\eqno(3.4) $$

Once this DR has been understood, it is straightforward to apply it to the
$~N=2$~ SKdV.  For example, we choose $~F_{A B}$~ to be {\it Abelian} in
the $~N=(2,0)$~ superspace with the super-coordinates $~(Z^M) =
(x,\,t,\,\theta^1,\,\theta^2)$, and
$$\eqalign{& \nabla_i \equiv D_i + D_i \Phi~~,  ~~~~
D_i \equiv \fracmm\partial {\partial \theta^i} + \theta^i \fracmm\partial
{\partial x}~~, ~~~{\scst (i,~j,~\cdots~=~ 1,~2)} ~~,  \cr
&\nabla_x \equiv \half \{ \nabla_1 ,\,\nabla_1\} =
\half \{ \nabla_2 ,\,\nabla_2\} = \fracmm\partial{\partial x} + A_x ~~,
{}~~~~\nabla_t \equiv \fracmm{\partial}{\partial t} + A_t ~~,
{}~~~~ F_{\theta^i \theta^j} = 0~~, \cr }
\eqno(3.5) $$
where the $~\Phi $~ is a {\it bosonic} superfield expanded as
$$\Phi(x,t,\theta^1,\theta^2)  = w(x,t) + \theta^1 \xi_1(x,t) + \theta^2
\xi_2 (x,t) + \theta^2 \theta^1 u(x,t) ~~.
\eqno(3.6) $$
As before, this superfield satisfies the usual boundary condition
$$\lim_{|x|\rightarrow\infty} \Phi(x,t,\theta^1,\theta^2) = 0
\eqno(3.7) $$
at any arbitrary $~t$.
Even though this condition seems rather artificial, it can be
interpreted as a sort of ``gauge-fixing''.  As a matter of fact, we see
that the system (3.3) - (3.5) is invariant under the ``gauge'' transformation
$$ \d \Phi = X(t) ~~,
\eqno(3.8) $$
where $~X(t)$~ is an arbitrary function {\it only} of $~t$, {\it independent
of} $~x$~ or $~\theta^i$, satisfying
$$ D_i X(t) = 0 ~~.
\eqno(3.9) $$
This interpretation is peculiar to the DR of the {\it second} type, which was
not clear in the first type.

It is easily seen that (3.5) implies
$$A_x = \Phi '~~,
\eqno(3.10) $$
while if we identify
$$A_t = - \Phi''' + 3 (\Phi D_1D_2 \Phi) ' + \half (a-1) (D_1D_2\Phi^2 )
' + 3a \Phi^2 \Phi ' ~~,
\eqno(3.11) $$
then $~F_{\theta^i t}$~ and $~F_{x t}$~ are
$$ \li{&F{\low{\theta^i t}} =
- D_i \left[ \Dot\Phi - \left( - \Phi ''' + 3 (\Phi D_1D_2 \Phi) '
+ \half (a-1) (D_1D_2\Phi^2 )' + 3a \Phi^2 \Phi ' \right)\right] ~~,
&(3.12) \cr
&F_{x t} = - \left[ \Dot\Phi - \left(- \Phi ''' + 3 (\Phi D_1D_2 \Phi) '
+ \half (a-1) (D_1D_2\Phi^2 )' + 3a \Phi^2 \Phi ' \right)\right] ' ~~.
&(3.13) \cr } $$
If we impose the condition $~F_{\theta^i t} = 0$, we get
$$\Dot\Phi = - \Phi ''' + 3 (\Phi D_1D_2 \Phi) '
+ \half (a-1) (D_1D_2\Phi^2 ) ' + 3a \Phi^2 \Phi ' + Y(t) ~~,
\eqno(3.14) $$
where $~Y(t)$~ is an arbitrary scalar function
{\it independent of} $~x$~ or $~\theta^i$, satisfying $~D_i
Y(t) = 0$.  However, such a function
is again excluded by our boundary condition (3.7), or to put it
differently, it is just a ``gauge'' freedom to be removed.  Eventually we get
the {\it unique} equation
$$\Dot\Phi = - \Phi ''' + 3 (\Phi D_1D_2 \Phi) '
+ \half (a-1) (D_1D_2\Phi^2 ) ' + 3a \Phi^2 \Phi ' ~~.
\eqno(3.15) $$
The constant $~a$~ is to be $~a=-2,~1$~ or $~4$~ for the equations
to be exactly soluble [9].  The only remaining condition $~F_{x t}=0$~ is now
automatically satisfied under (3.15).

As in the DR of the first type, we can obtain also the $~N=1$~ SKdV as
a sub-case by truncation.
Compared with the $~N=1$~ SKdV in the previous section, this
system is much simpler and its supersymmetry is manifest.  Even though other
components in the original field strength such as $~F_{y t}$~ {\it etc.}
disappear in the final system, they still satisfy the original SD condition
(2.1), as the most trivial solution.

The above system (3.3) has more applications.
The $~N=(1,1)$~ supersymmetric Liouville equation [15] is one of such good
examples.  We first put the supergauge potential [12,15]
$$ A_\theta = 2\b (D\Phi) L_0 + G_- ~~, ~~~~ A_{\Tilde\theta}
= - e^{\b\Phi} G_+ ~~,
\eqno(3.16) $$
where the generators $~L$'s and $~G$'s are those of the Lie superalgebra
$~OSp(1,2)$, satisfying
$$\eqalign{&\[ L_0\, , ~L_\pm \] = \mp L_\pm ~~, ~~~~
\[ L_+\, , ~L_-\] = 2 L_0 ~~, \cr
&\[ L_0\, , ~G_\pm \] = \mp \half G_\pm ~~, ~~~~
\[ L_\pm\, , ~G_\mp \] = \pm G_\pm ~~, \cr
&\{ G_+\, , ~ G_- \} = 2 L_0 ~~, ~~~~
\{ G_\pm \, , ~ G_\pm \} = 2L_\pm ~~. \cr}
\eqno(3.17) $$
The $~\Phi$~ is a scalar superfield, and $~D^2 = \partial_z,~ \Tilde D^2
= \partial_{\Tilde z},~ \{ D,\,\Tilde D\} = 0$.

Inserting these into the equation $~F_{\theta\Tilde\theta} = 0$, we can
immediately get the equation
$$F_{\theta\Tilde\theta} = -2 \left[\, \b (D \Tilde D \Phi) +
e^{\b\Phi}\,\right] L_0 = 0 ~~,
\eqno(3.18) $$
which yields nothing else than the usual supersymmetric Liouville equation
[15],
after identifying the grading in Lie superalgebra with the $~D=2$~
space-time supersymmetry as usual [12].  Other components $~F_{z
\theta} = 0,~ F_{z \Tilde\theta} = 0$~ and $~F_{\ula\ulb}=0$~ are easily
satisfied by
$$\eqalign{&\nabla_z \equiv \partial_z + A_z \equiv \nabla^2 ~~, ~~~~
\Tilde\nabla_{\Tilde z} \equiv \partial_{\Tilde z} + A_{\Tilde z} \equiv
\Tilde\nabla^2~~,   \cr
&\nabla \equiv D + A_\theta ~~, ~~~~ \Tilde\nabla \equiv \Tilde D
+ A_{\Tilde\theta} ~~. \cr }
\eqno(3.19) $$

The case of $~N=1$~ supersymmetric Toda theory works in a similar way
for what is called the almost simple contragradient Lie superalgebra
with pure fermionic root system [12].
Choose its Cartan-Weyl-type bases satisfying
$$\[H\, ,~H\] = 0~~, ~~~~ \[ H\, , ~e_i^\pm \] = \pm \a_i e_i^\pm ~~,
{}~~~~ \[ e_i^+ \, ,~ e_j^- \} = \d_{i j} \a_i\cdot H~~,
\eqno(3.20) $$
where $~\a_i$~ are as usual the simple roots, and $~H$~ is an
$~r\-$vector in the Cartan subalgebra of rank $~r$, which coincides with
the rank of the Cartan matrix $~a_{i j} = \a_i \cdot \a_j$.
The $~H$~ is always bosonic, while $~e_i^\pm$~ are either bosonic or
fermionic.  Now we can
introduce a real superfield $~\Phi$~ in the $~r\-$dimensional representation,
and put
$$\eqalign{&A_\theta = \b (D\Phi) \cdot H + \sum_{i \in {\cal F}}
e_i^+  ~~, \cr
& A_{\Tilde\theta} = - \sum_{i\in {\cal F}}
\exp(\b\a_i \cdot \Phi) e_i ^- ~~. \cr}
\eqno(3.21) $$
Insertion of these into the superfield equation $~F_{\theta\Tilde\theta} = 0$~
produces the $~N=1$~ supersymmetric Toda field equation [12]
$$D\Tilde D\Phi + \fracm1\b \sum_{i\in{\cal F}} \a_i \exp (\b \a_i \cdot
\Phi) = 0 ~~.
\eqno(3.22) $$
Here $~{\scst{\cal F}}$~ is for the {\it fermionic}
grading indices $~{\scst{\cal F} ~\subset ~ \{1,\,2,\,\cdots,\, r\} }$.
Other components in $~F_{A B}=0$~ also vanish under (3.19).

        We can take a similar procedure for $~N=2$~ supersymmetric
Toda field theory [12], but we skip it in this Letter, because its pattern is
essentially the same, while it costs more space for notational arrangements.

\bigskip\bigskip

\noindent 4.{\it ~~Concluding Remarks.}~~In this Letter we have shown that
the SDSYM theory in $~D=(2,2)$~ [2-5] can embed soluble systems in $~D=2$~
after appropriate DRs.  As typical examples, we gave the cases of
$~N=2$~ SKdV, $~N=1$~ supersymmetric Liouville equations and $~N=1$~
supersymmetric Toda gauge theories.  We gave the $~N=2$~ SKdV as an example
of our DR of the first type, while the others for the DR of the second type.

        The DR schemes we have given in this Letter are by no means the most
general ones.  We can perform an alternative scheme, where the $~A_x$~
and $~A_t$~ are {\it not} gauged away.  We stress that the examples we gave
form only a small subset of a much wider class of {\it supersymmetric}
soluble systems, that are generated by the $~D=(2,2)$~
SDSYM theories.  For example, we could have embedded the SKdV in the {\it
non-Abelian} $~SL(n)$~ group [5] in the SDSYM.

        The SKdV and supersymmetric Toda theories are just examples of our
embedding, and we
stress the importance of our results (2.9), (2.10), (2.14) - (2.16)
for the DR of the first type, which are most likely to have more
applications in the future, because they can potentially generate other
exactly soluble models.  We also emphasize the point we have made about the
superspace field equation $~F_{A B}=0$~ (3.3) that can embed the $~N=2$~ (and
$~N=1$~ as its sub-case) SKdV equations, which has never been presented in
the literature to our knowledge.

        We have started with the $~D=(2,2),\,N=2$~ SDSYM theory.
However, we can go further to the $~N=4$~ extended SDSYM
theories in $~D=(2,2)$, as constructed in our recent paper [4].  We can apply
similar DR rules, which create $~N=4$~ heterotic supersymmetric soluble
systems in $~D=2$.  It is also interesting to notice that the SDSG
theories are supposed to generate the $~W_\infty\-$algebra in $~D=2$~
[16].  The exact solutions for the coupled system of SDSYM + SDSG given
in a recent paper [17] may have interesting significance related to
compactifications.

        The master superfield equation $~F_{A B} =0$~ (3.3) indicates
a relation to topological field theory, and to the Chern-Simons theory [18]
in $~D=3$.  This is because the superfield strength vanishes, and the only
observables in such a system are super-Wilson loop integrals, e.g., for
the $~N=2$~ SKdV we have\footnotew{The combination $~(d x - d\theta^i
\theta_i)$~ is needed to comply with ~$d Z^M A_M = d Z^M E\du M B A_B$.}
$$\eqalign{&\exp\,\left[ \,i \oint d Z^M \, A_M \, \right] \cr
&~~ = \exp \, \Bigg[i\oint \bigg\{ (d x - d\theta^i\theta_i) \Phi '
+ d\theta^i D_i \Phi \cr
&~~~~~ ~~~~~ ~~~~~ ~~~~~+ \Big( - \Phi''' + 3(\Phi D_1D_2 \Phi)' + \half (a-1)
(D_1 D_2 \Phi^2)' + 3a \Phi^2 \Phi' \Big)  d t \bigg\}\Bigg] ~~. \cr}
\eqno(4.1) $$
If the superfield equation (3.15) is used, the integrand is {\it exact}:
$~\oint d\Phi$.
Moreover, a simple consideration of DR gives the link between
$~D=3$~ supersymmetric Chern-Simons (SCS) theory [19] and the $~D=4$~
SDSYM.  These viewpoints strongly support the idea of a close relationship
between the $~D=2$~ supersymmetric soluble systems and the $~D=3$~ SCS theory.

        Once our DR rules have been established, there are many applications.
One good example is to seek the corresponding four-dimensional significance
for the conserved charges in two-dimensional soluble systems [6-8,10].
The fact that the $~N=2$~ SKdV system is described by {\it pure
gauge} superfield strength suggests the close relationship between
supersymmetric integrability in $~D=2$~ and supersymmetric SD in $~D=4$.

        The $~N=2$~ SDSYM theory is the {\it
maximal one} in $~D=(2,2)$~ due to the barrier described in Ref.~[5]
beyond $~N=2$, {\it unless} we introduce propagating multiplier
fields [5,20].  It is likely that there is also a similar barrier for
extended supersymmetric soluble systems in $~D=2$.  In the case of SDSG,
the barrier exists between $~N=4$~ and larger $N$.  We can find a good
example about corresponding barriers (after DR) in the $~D=2$~ extended
superconformal theories,
that exists only up to $~N=4$~ due to the involvement of conformal
compensators [21], indicating the {\it obstruction} for finding soluble
(free field) systems beyond $~N=4$~ in $~D=2$.

\vfill\eject

\refs
\normalsize

\items{1} H. Ooguri and C. Vafa, \mpl{5}{90}{1389};
\np{361}{91}{469}; \ibid{367}{91}{83};
H.~Nishino and S.J.~Gates, Jr., Maryland preprint,
UMDEPP 92--137 (Jan.~1992), to appear in Mod.~Phys.~Lett.

\items{2} S.V.~Ketov, S.J.~Gates and H.~Nishino, Maryland preprint,
UMDEPP 92--163 (February 1992).

\items{3} H. Nishino, S. J. Gates, Jr. and S. V. Ketov,
Maryland preprint, UMDEPP 92--171 (February 1992).

\items{4} S. J. Gates, Jr., H. Nishino and S. V. Ketov, Maryland
preprint, UMDEPP 92--187 (March 1992), to appear in Phys.~Lett.B.

\items{5} S.J.~Ketov, H.~Nishino and S.J.~Gates, Jr., Maryland preprint,
UMDEPP 92--211 (June 1992), to appear in Nucl.~Phys.~B.

\items{6} A.A.~Belavin, A. M. Polyakov, A. Schwartz and Y. Tyupkin,
\pl{59}{75}{85};
R.S.~Ward, \pl{61}{77}{81};
M.F.~Atiyah and R.S.~Ward, \cmp{55}{77}{117};
E.F.~Corrigan, D.B.~Fairlie, R.C.~Yates and P.~Goddard,
\cmp{58}{78}{223};
E.~Witten, \prl{38}{77}{121};
A.N.~Leznov and M.V.~Saveliev, \cmp{74}{80}{111};
L.~Mason and G.~Sparling, \pl{137}{89}{29};
I.~Bakas and D.A.~Depireux, \mpl{6}{91}{399}; {\it ibid.} 1561;
2351.

\items{7} M. F. Atiyah, unpublished;
R. S. Ward, Phil.~Trans.~Roy.~Lond.~{\bf A315} (1985) 451;
N. J. Hitchin, Proc.~Lond.~Math.~Soc.~{\bf 55} (1987) 59.

\items{8} G.~Segal and G.~Wilson, Publ.~Math.~~IHES {\bf 61} (1985) 5;
V.G.~Drinfeld and V.V.~Sokolov, Sov.~Math.~Dokl.~{\bf 23}
(1981) 457; Jour.~Sov.~Math.~{\bf 30} (1985) 1975;
For a review see Yu.I.~Manin, J.~Sov.~Math.~{\bf 11} (1979) 1.

\items{9} Yu.I.~Manin and \& A.O.~Radul, \cmp{98}{85}{65};
P.~Mathieu, Jour.~Math.
Phys.~{\bf 29} (1988) 2499;
C.A.~Laberge and P.~Mathieu, \pl{215}{88}{718};
T.~Inami and H.~Kanno, YITP-Kyoto preprint YITP/K-929 (July 1991).

\items{10} l.J.~Mason and G.A.J.~Sparling, Phys.~Lett.~{\bf 137B} (1989)
29;
L.J.~Mason, Twist.~Newslett.~{\bf 30} (1990) 14;
I.~Bakas and D.~Depireux, Maryland preprints, UMDEPP 91--088
UMD-PP 91-111 (Nov.~1991).

\items{11} A.V.~Mijailov, JETP Lett.~{\bf 30} (1980) 414;
A.N.~Lezanov and M.V.~Saveliev, Lett.~Math.~Phys.
{\bf 3} (1979) 489;
C.A.~Bulgadaev, \pl{96}{80}{151};
V.A.~Fateev and A.B.~Zamolodchikov, \ijmp{5}{90}{1025};
T.~Eguchi and S.K.~Yang, \pl{224}{89}{373};
T.J.~Hollowood and P.~Mansfield, \pl{226}{89}{73}.

\items{12} L.A.~Leites, M.V.~Saveliev and V.V.~Serpukgov, ``Embeddings of
of $~OSp(N/2)$~ and Associated Non-linear Supersymmetric Equations'',
{\it in} Proc.~Third Yurmala Seminar (USSR 22-24 May 1985);
M.A.~Olshanetsky, \cmp{88}{83}{63};
J.~Evans and T.~Hollowood, \np{352}{91}{723}.

\items{13} J.~Scherk and J.H.~Schwarz, \np{153}{79}{61}.

\items{14} A.~Das and C.A.P.~Galv{\~ a}o, Rochester preprint,
UR-1274, ER-40425-260 (1992).

\items{15} See, e.g., J.F.~Arris, \np{212}{83}{151}, \ibid{218B}{83}{309};
H.C.~Liao and P.~Mansfield, \np{344}{90}{696}.

\items{16} Q.-H.~Park, Cambridge preprint, DAMTP R-91/12 (Oct.~1991).

\items{17} H.~Nishino, Maryland preprint, UMDEPP 93--52 (Sept.~1992).

\items{18} A.~Schwarz, Lett.~Math.~Phys.~{\bf 2} (1978) 247;
W.~Siegel, \np{156}{79}{135};
J.~Schonfeld, \np{185}{81}{157}; R.~Jackiw and S.~Templeton,
\pr{23}{81}{2291}; C.R.~Hagen, \ap{157}{84}{342}; \pr{31}{85}{331};
E. Witten, \cmp{121}{89}{351}.

\items{19} S.J.~Gates, M.T.~Grisaru, M.~Ro{\v c}ek and W.~Siegel, ``{\it
Superspace}'', (Benjamin/Cummings, Reading MA, 1983), page~27;
H.~Nishino and S.J.~Gates, Maryland preprint, UMDEPP 92--060, to appear
in Int.~Jour.~Mod.~Phys.

\items{20} W.~Siegel, Stony Brook preprint, ITP-SB-92-24 (May, 1992).

\items{21} E.~Bergshoeff, H.~Nishino and E.~Sezgin, \pl{185}{87}{167}.

\end{document}